\documentclass[12pt]{article}
\usepackage{amssymb}  
\usepackage{amsthm}
\usepackage{amsmath}
\usepackage{mathrsfs}
\usepackage{setspace}
\usepackage{bm}
\begin{document}

\begin{titlepage}

\begin{flushright}
Imperial/TP/16/AH/01\\
QMUL-PH-15-27\\
\end{flushright}
\vskip 2cm

\begin{center}

\begin{spacing}{2}
{\Large \bfseries 
Highest Weight Generating functions for hyperK\"ahler $T^{\star}(G/H)$ spaces}
\end{spacing}

\vskip 1.2cm

Amihay Hanany$^{a}$\footnote{a.hanany@imperial.ac.uk},  Sanjaye Ramgoolam$^{b}$\footnote{s.ramgoolam@qmul.ac.uk} and Diego Rodriguez-Gomez$^{c}$\footnote{d.rodriguez.gomez@uniovi.es}

\bigskip
\bigskip

\begin{tabular}{c}
$^{a}$ Theoretical Physics Group, Imperial College London, \\
Prince Consort Road, London, SW7 2AZ, UK\\ 
\\
$^{b}$ Centre for Research in String Theory, School of Physics and Astronomy\\
Queen Mary University of London\\    Mile End Road, London E1 4NS, UK\\
\\
$^{c}$ Department of Physics, Universidad de Oviedo, \\
Avda.~Calvo Sotelo 18, 33007, Oviedo, Spain
\end{tabular}

\vskip 1.5cm

\textbf{Abstract}
\end{center}

\medskip
\noindent
We develop an efficient procedure for counting  holomorphic functions on a hyperKahler cone that has a resolution as a cotangent bundle of a homogeneous space by providing a formula for computing the corresponding Highest Weight Generating function.

\bigskip
\vfill
\end{titlepage}

\setcounter{tocdepth}{2}

\section{Introduction}

HyperK\"ahler cotangent bundles over a homogeneous space,  $T^{\star}(G/H)$,  arise 
in a number of contexts in supersymmetric gauge theories, brane physics, and in geometry.  For instance, they bear a direct relation to solutions of the Nahm equations \cite{K} and hence to moduli spaces of $T^{\sigma}(G)$ theories \cite{Gaiotto:2008sa,Gaiotto:2008ak}. More generically, these spaces appear as building blocks of theories with 8 supercharges (in particular stemming from \cite{Gaiotto:2009we}. See also \cite{Nanopoulos:2009uw}), and have interesting implications for $\mathcal{N}=1^{\star}$ theories as well \cite{Bourget:2015cza}. In fact, they are intimately related to a subject recently blossoming in a number of different contexts in String Theory, namely the theory of nilpotent orbits (for an introduction, see \cite{C-MG}). 

Their avatars as Coulomb branches of $T^{\sigma}(G)$ theories, available for classical groups, is particularly interesting. For instance, concentrating on the case of $G=SU(N)$, $\sigma$ is a partition of $N$ specifying the brane system \cite{Hanany:1996ie} that realizes the theory, which is a linear quiver with flavors for each node. Moreover, it specifies in a prescribed way a Levi subgroup $H$ of $G$ such that the Coulomb branch moduli space is $T^{\star}(G/H)$ (we refer to \cite{Gaiotto:2008sa,Gaiotto:2008ak,Chacaltana:2012zy}. Note that this bears also interesting relations to 3d indices through the Coulomb branch formula \cite{Cremonesi:2013lqa,Razamat:2014pta,Cremonesi:2014uva}). Other appearances of these spaces are in the context of instanton moduli spaces, including also the cases of exceptional Lie algebras (see \textit{e.g} \cite{Cremonesi:2014xha}).

On general grounds, an object of primary interest on a complex variety is the ring of (polynomial) holomorphic functions defined on it. This object carries a great deal of information about the underlying variety. A particularly efficient device to encode its properties is the so-called Highest Weight Generating function introduced in \cite{Hanany:2014dia}. Our goal in this note is to provide a very elegant and efficient way to compute the (unrefined, \textit{i.e.} $t=1$) Highest Weight Generating function for hyperK\"ahler cotangent bundles over homogeneous spaces.

In previous approaches one realizes the space $T^{\star}(G/H)$ as the vacuum moduli space on either the Higgs or Coulomb branch of some (3d) gauge theory, compute the Hilbert series either by the Molien integral (see \textit{e.g} \cite{Benvenuti:2010pq}) or the monopole formula (akin to an index) and then read off the Highest Weight Generating function as in \cite{Cremonesi:2014uva}. Instead, in this note we propose a much shorter path stemming from the observation that the functions in $T^{\star}(G/H)$ are associated to representations of $G$ containing singlets of $H$ when branched under $H$. 

The rest of this note is organized as follows: in section \eqref{HWG} we provide, for completeness, a lightning review of Highest Weight Generating functions. In section \eqref{conjecture} we describe precisely our conjecture -- explicitly captured by eq.\eqref{master_formula} -- for which we offer examples and tests in section \eqref{examples}. We conclude in section \eqref{conclusions} with some open problems.

\section{Highest Weight Generating functions}\label{HWG}

On general grounds, ennumerating holomorphic functions on a complex variety $\mathcal{M}$ is of great interest. Typically, these functions are labelled by their quantum numbers under the isotropy group -- which plays the role of a global symmetry -- and graded in a certain way.\footnote{Even though for our purposes it will not play an essential role, the grading corresponds to the highest weight (twice the spin) of the $SU(2)_R$ representation. In more physical terms, the spaces at hand can be thought as moduli spaces of theories with 8 supercharges. In such theories the R-symmetry contains an $SU(2)$ which acts antiholomorphically on a hypermultiplet. Thus, in 4 supercharge language, only its highest weight $r$ is visible, assigning -- in a certain normalization -- $r= 1$ to both complex fields in a hypermultiplet. This highest weight can be used to grade chiral operators -- a.k.a. holomorphic functions on the moduli space. Note that $r$ is proportional to the scaling dimension $\Delta$ of the operators.}  More precisely, introducing fugacities $\mathbf{z}$ for the global symmetry and $t$ for the grading, each function with global charges $\mathbf{q}$ and corresponding grading $r(\mathbf{q})$ can be encoded in a monomial $\mathbf{z}^{\mathbf{q}}t^{r(\mathbf{q})}$. The sum of these monomials is a generating function, counting holomorphic functions in the variety, called Hilbert series\footnote{Note that more than one function might have the same quantum numbers. In that case the corresponding monomial will appear as many times as functions with those quantum numbers, that is, it will have some non-trivial multiplicity $m(\mathbf{q})$.}

\begin{equation}
HS[\mathcal{M}](t;\mathbf{z})=\sum_{\mathbf{q}}\,m(\mathbf{q})\mathbf{z}^{\mathbf{q}}t^{r(\mathbf{q})}\,.
\end{equation}

The coefficients of $t$ in the expansion of the Hilbert series group into characters of the global symmetry, reflecting the fact that holomorphic functions form multiplets of such global symmetry. Since, in turn, each representation can be labelled by the highest weight state with Dynkin labels $\mathbf{n}$ -- that is, we encode in the entries of the vector $\mathbf{n}$ the Dynkin labels of the representation $[n_1,\cdots,n_{{\rm rank}G}]$ -- in \cite{Hanany:2014dia} a more concise encoding of the same information was introduced through the so called Highest Weight Generating function (HWG). The idea is to introduce a set of fugacities $\bm{\mu}$ so that the whole multiplet associated to the highest weight $\mathbf{n}$ contributes $\bm{\mu}^{\mathbf{n}}\,t^{r(\mathbf{n})}$. Here $\bm{\mu}^{\mathbf{n}}=\mu_1^{n_1}\cdots \mu_{{\rm rank}\,G}^{n_{{\rm rank}\,G}}$. Then, the HWG is (again, non-trivial multiplicities $\hat{m}(\mathbf{n})$ may appear)

\begin{equation}
g^{\mathcal{M}}(t;\bm{\mu})=\sum_{\mathbf{n}}\hat{m}(\mathbf{n})\bm{\mu}^{\mathbf{n}}\,t^{r(\mathbf{n})}\, .
\end{equation}

Let us put this into practice with the simple example of the moduli space of one $SU(2)$ instanton, which corresponds to the minimal nilpotent orbit of $SU(2)$. Stripping off the center of mass, the Hilbert series is just that of $\mathbb{C}^2/\mathbb{Z}_2$ \cite{Benvenuti:2010pq}

\begin{equation}
HS\left[\mathbb{C}^2/\mathbb{Z}_2\right](t;z)=\frac{1+t^2}{(1-t^2z^2)\,(1-t^2z^{-2})}\, ,
\end{equation}
where $z$ is the fugacity for the global $SU(2)$ symmetry that commutes with $SU(2)_R$ associated to the $\mathbb{C}^2/\mathbb{Z}_2$ space. Expanding this we have

\begin{equation}
HS\left[\mathbb{C}^2/\mathbb{Z}_2\right](t;z)=1+\chi[2]\,t^2+\chi[4]\,t^4+\cdots\, ,
\end{equation}
where $\chi[n]$ represents the character of the $[n]$ representation (of dimension $n+1$) of $SU(2)$ in terms of the fugacity $z$. Thus, we see that only $[2n]$ appears. Hence, the HWG is

\begin{equation}
g^{\mathbb{C}^2/\mathbb{Z}_2}(t;\mu_1)=\sum_{n=0}^\infty \mu_1^{2n}\,t^{2n}=\frac{1}{1-\mu_1^2\,t^2}\, .
\end{equation}
Upon setting $t=1$ we find the unrefined HWG, which in this case reduces to $g^{\mathbb{C}^2/\mathbb{Z}_2}(\mu_1)={\rm PE}[\mu_1^2]$.

Another device which will be useful for our purposes below is the so-called character generating function for a certain group $G$. This function, also introduced in \cite{Hanany:2014dia}, is designed so that the coefficient of the $\bm{\mu}^{\mathbf{n}}$ term in its expansion gives the character $\chi[\mathbf{n}]$ for the representation of $G$ whose Dynkin labels are $\mathbf{n}$. In the following, we will use as group fugacities $\mathbf{z}$, $\mathbf{w}$ and $u$, and thus we will denote character generating functions as \textit{e.g.} ${\rm g}^{\rm G}(\bm{\mu};\mathbf{z})$.

Let us make this precise with the $SU(2)$ example. The character of the $SU(2)$ representation with Dynkin label $[n]$ is given by

\begin{equation}
\chi[n]=\frac{z^{n+1}-z^{-(n+1)}}{z-z^{-1}}\, .
\end{equation}
Thus, the character generating formula for $SU(2)$ representations is simply

\begin{equation}
\label{chi_SU(2)}
{\rm g}^{SU(2)}(\mu_1;z)=\sum_{n_1=0}^{\infty} \chi[n_1]\mu_1^{n_1}=\frac{1}{(1-\mu_1\,z)\,(1-\mu_1\,z^{-1})}\, ,
\end{equation}
in such a way that the coefficient of $\mu_1^n$ in the expansion of $g^{SU(2)}(\mu_1,\,z)$ gives the character of the $[n]$ representation of $SU(2)$. 

Note that this can be extended in a straightforward way to $SU(N)$, since the Weyl character formula allows to write $U(N)$ characters in terms of Young tableux as

\begin{equation}
\chi[\mathbf{n}]=\frac{{\rm det}(\hat{z}_i^{\hat{r}_j+N-j})}{{\rm det}(\hat{z}_i^{N-j})}\, ,\qquad i,\,j=1\cdots N\, .
\end{equation}
Projecting to $SU(N)$ is done by setting $\hat{z}_1=z_1,\hat{z}_2=\frac{z_2}{z_1},\cdots,\hat{z}_N=\frac{1}{z_{N-1}}$. This allows to set $\hat{r}_1=r_1+\hat{r}_N,\cdots, \hat{r}_{N-1}=r_{N_1}+\hat{r}_N$ in such a way that $\hat{r}_N$ drops from the formula. Then, the character generating function for $SU(N)$ is

\begin{equation}
\label{chi_SU(N)}
{\rm g}^{SU(N)}(\bm{\mu};\mathbf{z})=\sum_{r_1=0}^{\infty}\sum_{r_2=0}^{r_1}\cdots\sum_{r_{N_1}=0}^{r_{N-2}}\,\chi[\mathbf{n}]\,\bm{\mu}^{\mathbf{n}}\,.
\end{equation}
It is straightforward to see that this formula reproduces \eqref{chi_SU(2)} in the $N=2$ case.

\section{HWG for hyperK\"ahler $T^{\star}(G/H)$}\label{conjecture}

As discussed in the introduction, hyperK\"ahler spaces of the form $T^{\star}(G/H)$ are very interesting for a number of reasons. On general grounds, at least locally, the cotangent bundle over $G/H$ is hyperK\"ahler if $G/H$ is K\"ahler. On the other hand, $G/H$ is K\"ahler if ${\rm rank}(G)={\rm rank}(H)$ and $H=H'\times U(1)$ (see \textit{e.g.} \cite{Ketov:1995yd,Higashijima:2002px}). Thus, we will restrict to those cases. Note that, in particular, for $G=SU(N)$ the possible $H$ are in one-to-one correspondence with partitions of $N$. Note also that the condition ${\rm rank}(G)={\rm rank}(H)$ is needed to ensure that the characteristic polynomial of the adjoint valued generators of $T^{\star}(G/H)$ vanishes (this is equivalent to the vanishing of all Casimir invariants, hence the name nilpotent). In turn, this has to be the case, as these spaces arise in particular as moduli spaces of $T^{\sigma}(G)$ theories and thus correspond to nilpotent orbits.

A key fact we will use is that the hyperK\"ahler manifold $T^{\star}(G/H)$ is equivalently realized as  $ G^{\mathbb{C}}/H^{\mathbb{C}} $
\cite{Arai-Nitta,Kron-hyp-coad}.  Our primary interest  is in the counting of the set of holomorphic functions on this space.
 As discussed above, a particularly convenient way to encode such functions is the HWG. Thus, we will be interested on the ($t$-unrefined) HWG for $T^{\star}(G/H)$. 
The realization as  $G^{\mathbb{C}}/H^{\mathbb{C}}$ allows us to compute the HWG function using a standard result 
from the theory of homegeneous spaces: namely that the functions can be decomposed into representations of $G$ which contain one or more singlets of $H$. 
The multiplicity of singlets leads to a multiplicity of the $G$-representations \cite{JHYang}. This is a consequence of the Peter-Weyl theorem  

We can motivate our procedure from physical intuition, as we might think of $G^{\mathbb{C}}/H^{\mathbb{C}}$ as the target space manifold of a low energy effective theory after symmetry breaking (see \cite{Lerche:1983qa} and \cite{Nitta:1998qp} for a recent analysis). In this context, the operation of keeping representations of $G$ containing singlets under $H$ naturally counts chiral operators in this effective theory and hence holomorphic functions on the target space $T^{\star}(G/H)$.

\section{Examples}\label{examples}

Since we have proposed that holomorphic functions on $T^{\star}(G/H)$ correspond to representations of $G$ containing $H$ singlets, we have an operationally easy procedure to construct and ennumerate all such functions -- that is, to construct the HWG. Let us put this into practice with examples. In principle, we can compute the HWG of the cotangent bundle over $G/H$ by brute force decomposing the representations of $G$ under $H$ and selecting by hand those containing singlets of $H$. Then, re-summing the series we can obtain the HWG at $t=1$. However, a more refined approach is to start with the character generating function of $G$. Then, the projection to representations containing $H$-singlets is tantamount to gauging $H$. Thus, integration over $H$ of the character generating function of $G$ will precisely pick the representations containing singlets, labelling them by their highest weight. Thus, the HWG at $t=1$ of $T^{\star}(G/H)$ can be easily computed as

\begin{equation}
\label{master_formula}
g^{T^{\star}(G/H)}=\int d\mu_H\,{\rm g}^{G}\, ,
\end{equation}
where $\int d\mu_H$ represents the integration over $H$ including its Haar measure.

\subsection{$G=SU(2)$}\label{SU(2)/U(1)}

The character generating function for $SU(2)$ is shown in \eqref{chi_SU(2)}. On the other hand, for $SU(2)$, the subgroup $H$ with ${\rm rank}(H)={\rm rank}(SU(2))$ can only be $H=U(1)$. Note that $T^{\star}(SU(2)/U(1))$ is the resolution of $\mathbb{C}^2/\mathbb{Z}_2$, and therefore we should recover the results above.  Moreover, it is clear that $H=U(1)$ is precisely the Cartan of $SU(2)$, and thus its fugacity is simply $z$. Hence, we can easily project $g^{SU(2)}$ down to $H$-singlets to find $g^{T^{\star}(SU(2)/U(1))}$ as

\begin{equation}
g^{T^{\star}(SU(2)/U(1))}(\mu_1)=\int\frac{dz}{z}\,{\rm g}^{SU(2)}(\mu_1,\,z)=\frac{1}{1-\mu_1^2}=\sum_{n=0}^{\infty}\,\mu_1^{2n}\, .
\end{equation}
Thus we see that in this case only the reps $[2n]$ survive the projection, exactly as expected for $T^{\star}(SU(2)/U(1))$ corresponding to $\mathbb{C}^2/\mathbb{Z}_2$. Moreover we have that $g^{T^{\star}(SU(2)/U(1))}={\rm PE}[\mu_1^2]$, and we therefore see that the generator of the holomorphic functions on $T^{\star}(SU(2)/U(1))$ is the adjoint of $SU(2)$.

Note that the spaces $T^{\star}(SU(N)/U(N-1))$ can be regarded as reduced moduli spaces of one instanton of $SU(N)$, whose corresponding HWG have been computed in \cite{Hanany:2014dia,Hanany:2015hxa}. The result above reassuringly matches the expected one.

\subsection{$G=SU(3)$}

Using \eqref{chi_SU(N)}, the $SU(3)$ character generating function is

\begin{equation}
\label{HGW_SU(3)}
{\rm g}^{SU(3)}(\mathbf{t};\mathbf{z})=\frac{1-\mu_1\,\mu_2}{(1-\mu_1\, z_1)\,(1-\frac{\mu_1}{ z_2})\,(1-\mu_1\,\frac{z_2}{z_1})\,(1-\mu_2\,z_2)\,(1-\frac{\mu_2}{z_1})\,(1-\mu_2\,\frac{z_1}{z_2})}\, .
\end{equation}

In the case of $SU(3)$ there are two possible $H$, namely $U(1)^2$ and $U(2)$. Let us treat both separately

\subsubsection{$H=U(1)^2$}

In this case $T^{\star}\left(SU(3)/U(1)^2\right)$ corresponds to the maximal nilpotent orbit of $SL(3,\mathbb{C})$ of complex dimension 6.
The $U(1)^2$ simply is the Cartan subalgebra of $SU(3)$. Hence the corresponding fugacities are directly the $\mathbf{z}$. Thus, we can compute the HWG of $g^{T^{\star}(SU(3)/U(1)^2)}$ as

\begin{eqnarray}
g^{T^{\star}(SU(3)/U(1)^2)}(\bm{\mu})&=&\int \frac{dz_1}{z_1}\int \frac{dz_2}{z_2}\,{\rm g}^{SU(3)}(\bm{\mu};\mathbf{z})=\\ \nonumber && \frac{1+\mu_1\mu_2+\mu_1^2\mu_2^2}{(1-\mu_1)\,(1+\mu_1+\mu_1^2)\,(1-\mu_2)\,(1+\mu_2+\mu_2^2)\,(1-\mu_1\mu_2)}\, .
\end{eqnarray}
This can be neatly written as

\begin{equation}
g^{T^{\star}(SU(3)/U(1)^2)}(\bm{\mu})={\rm PE}\left [\mu_1^3+\mu_2^3+2\,\mu_1\mu_2-\mu_1^3\mu_2^3\right]\, .
\end{equation}
One can check that this result is indeed consistent with the Hilbert series for the corresponding $T^{\sigma}(SU(3))$ theory as computed in \cite{Cremonesi:2014uva}.

\subsubsection{$H=U(2)$}\label{SU(3)/U(2)}

In this case $T^{\star}(SU(3)/U(2))$ corresponds to the minimal nilpotent orbit of $SL(3,\mathbb{C})$ of complex dimension 4.
Writing $z_1=u\,w$, $z_2=u^{-1}\,w$, the character of the fundamental representation arising from \eqref{HGW_SU(3)} becomes $u^{-2}+u\,(w+w^{-1})$. In these coordinates, it is clear that $H=U(2)=U(1)\times SU(2)$ is parametrized by $u$ for the $U(1)$ charges and $w$ for $SU(2)$. Thus, projecting to $H$-singlets yields the HWG for $T^{\star}(SU(3)/U(2))$

\begin{equation}
g^{T^{\star}(SU(3)/U(2))}(\bm{\mu})=\int\frac{du}{u}\int dw\,\frac{1-w^2}{w}\,{\rm g}^{SU(3)}(\bm{\mu};u,w)= \frac{1}{1-\mu_1\mu_2}\, .
\end{equation}
This can be re-written as

\begin{equation}
g^{T^{\star}(SU(3)/U(2))}(\bm{\mu})={\rm PE}[\mu_1\mu_2]\, ,
\end{equation}
which reproduces the expected result as in \cite{Hanany:2014dia}. It again shows the adjoint representation as generator of the space. 

\subsection{$G=SU(4)$}

Making use of \eqref{chi_SU(N)}, it is possible to resum the expression and explicitly compute the HWG for $SU(4)$. Yet, its form is very cumbersome and we will refrain from explicitly quoting it (see nevertheless \cite{Hanany:2014dia}). Moreover, in this case the possible $H$ are $U(1)^3$, $U(2)\times U(1)$, $S(U(2)\times U(2))$ and $U(3)$. For simplicity we will only focus on the last two cases.

\subsubsection{$H=S(U(2)\times U(2))$}\label{SU(4)/U(2)xU(2)}

In this case $T^{\star}(SU(4)/S(U(2)\times U(2)))$ corresponds to the next to minimal nilpotent orbit of $SL(4,\mathbb{C})$ of complex dimension 8.
Writing $z_1=u\,w_{1}$, $z_2=u^2$ and $z_3=u\,w_{2}$ the character of the fundamental of $SU(4)$ becomes $u(w_{1}+w_{1}^{-1})+u^{-1}(w_{2}+w_{2}^{-1})$. Thus we see that $w_{i}$ parametrizes each of the $SU(2)$'s inside $S(U(2)\times U(2))$, while $u$ parametrizes the antidiagonal combination of the $U(1)$'s which remains to form $S(U(2)\times U(2))$. Thus, the HWG for $T^{\star}(SU(4)/S(U(2)\times U(2)))$ is

\begin{eqnarray}
g^{T^{\star}(SU(4)/S(U(2)\times U(2)))}(\bm{\mu})&=&\int \frac{du}{u}\,\prod_{i=1}^2 \int dw_{i}\frac{1-w_{i}^2}{w_{i}}\,{\rm g}^{SU(4)}(\bm{\mu};u,w_{i})= \nonumber \\ && \frac{1}{(1-\mu_2^2)\,(1-\mu_1\,\mu_3)}\, .
\end{eqnarray}
This can be re-written as

\begin{equation}
g^{T^{\star}(SU(4)/S(U(2)\times U(2)))}(\bm{\mu})={\rm PE}[\mu_2^2+\mu_1\mu_3]\,.
\end{equation}
Again, it is easy to check that this result is consistent with the Hilbert series for the corresponding $T^{\sigma}(SU(4))$ theory computed in \cite{Cremonesi:2014uva}.

In fact this example can be regarded as part of the general family $T^{\star}(U(N)/U(N-k)\times U(k))$. This space appears as the Higgs branch moduli space of SQCD with 8 supercharges for gauge group $U(k)$ and $N$ flavors. As another example, it is easy to check that the $SU(5)/S(U(3)\times U(2))$ case has as HWG

\begin{equation}
g^{T^{\star}(SU(5)/S(U(3)\times U(2)))}(\bm{\mu})={\rm PE}[\mu_2\mu_3+\mu_1\mu_4]\,.
\end{equation}
In general, provided that $N\geq 2k$, one can convince oneself \cite{Gray:2008yu, Benvenuti:2010pq} that the HWG is

\begin{equation}
\label{HWG_SQCD}
g^{T^{\star}(U(N)/U(N-k)\times U(k))}(\bm{\mu})={\rm PE}\left[\sum_{i=1}^k\mu_i\,\mu_{N-i}\right]\, .
\end{equation}
The cases $N=2k$ for $k=1$ and $k=2$ are presented respectively in sections \eqref{SU(2)/U(1)} and \eqref{SU(4)/U(2)xU(2)} respectively. 

\subsubsection{$H=U(3)$}\label{SU(4)/U(3)}

In this case $T^{\star}(SU(4)/U(3))$ corresponds to the minimal nilpotent orbit of $SL(4,\mathbb{C})$ of complex dimension 6.
We write $z_1=u\,w_1$, $z_2=u^{-2}\,w_1$ and $z_3=u^{-1}w_1w_2^{-1}$, we can explicitly see $H=U(3)=U(1)\times SU(3)$, where $u$ parametrizes $U(1)$ and $\mathbf{w}$ parametrizes the $SU(3)$. Then, the HWG for $T^{\star}(SU(4)/U(3))$ is

\begin{eqnarray}
g^{T^{\star}(SU(4)/U(3))}(\bm{\mu})&=&\int\frac{du}{u}\int\frac{dw_1}{w_1}\frac{dw_2}{w_2}(1-w_1w_2)\,(1-\frac{w_1^2}{w_2})\,(1-\frac{w_2^2}{w_1})\,{\rm g}^{SU(4)}(\mathbf{t};u,\mathbf{w})\nonumber \\ =&& \frac{1}{1-\mu_1\mu_3}\,.
\end{eqnarray}
This can be re-written as

\begin{equation}
g^{T^{\star}(SU(4)/U(3))}(\bm{\mu})={\rm PE}[\mu_1\mu_3]\, ,
\end{equation}
which shows the adjoint as the generator and coincides with the expected result \cite{Hanany:2015hxa}. In fact, in view of the $T^{\star}(SU(2)/U(1))$, $T^{\star}(SU(3)/U(2))$ and $T^{\star}(SU(4)/U(3))$ cases, and setting $k=1$ in \eqref{HWG_SQCD}, we can conjecture the general form \cite{Benvenuti:2010pq} for $T^{\star}(SU(N)/U(N-1))$

\begin{equation}
g^{T^{\star}(SU(N)/U(N-1))}(\bm{\mu})={\rm PE}[\mu_1\mu_{N-1}]\, ,
\end{equation}
which corresponds to a space generated by the adjoint representation. It is straightforward to check that also the $N=5$ follows this prescription. Note that this space, which corresponds to the minimal nilpotent orbit of $SU(N)$, is the reduced one-instanton moduli space of $SU(N)$, and hence has complex dimension $2N-2$.

\subsection{Next to minimal orbit of $E_6$}

We now consider $G=E_6$ and $H=SO(10)\times U(1)$. This case corresponds to the so-called next to minimal nipotent orbit of $E_6$ of complex dimension 32. In this case following the above methods is hopeless, as finding the HWG for $E_6$ is very complicated. However, by explicitly branching the first few representations of $E_6$ into $SO(10)\times U(1)$ using \verb+LieART+ \cite{LieART}, and selecting those containing singlets of $SO(10)\times U(1)$ one can convince oneself that the HWG for $g^{T^{\star}(E_6/SO(10)\times U(1))}$ is

\begin{equation} 
g^{T^{\star}(E_6/SO(10)\times U(1))}(\bm{\mu})={\rm PE}[\mu_2+\mu_1\mu_6]\,.
\end{equation}
This is again consistent with expectations \cite{HK} (see also \cite{Cremonesi:2014uva}).

\section{Conclusions and open directions}\label{conclusions}

HyperK\"ahler spaces of the form $T^{\star}(G/H)$ are very interesting, as they appear in a number of situations of relevance in physics: building blocks of gauge theories with 8 supercharges, instanton moduli spaces, non-linear $\sigma$-model target spaces; to name just a few. In this note we have provided a very simple method to compute HWG for hyperK\"ahler spaces of the form $T^{\star}(G/H)$. This provides an efficient way to list the global charges of the holomorphic functions -- a.k.a. chiral operators in the physical language -- on these spaces. While we have checked our formula against a number of examples (of which we displayed only a subset to ease the presentation), it will be  very interesting to fully clarify the origin of this formula.

Nilpotent orbits are a class of co-adjoint orbits, obtained when a Lie group $ G $ acts on the dual $\mathfrak{g}^{\star}$ 
of its Lie algebra $\mathfrak{g}$. In this paper we  have exploited relations between the co-adjoint orbits of compact groups such as 
$SU(N)$ and their complexifications such as  $ SL(N , \mathbb{C} )$. This has been used along with the properties of functions on 
 coset spaces  $G^{\mathbb{C}}/H^{\mathbb{C}}$ to give a simple rule for the $G^{\mathbb{C}}$ representation content (Highest Weight Generating functions) 
 of hyperKahler spaces $ T^* ( G / H )$, demonstrating agreement with previous computations based on Higgs and Coulomb branches of quiver gauge theories. 
 Co-adjoint orbits have  been extensively studied in the context of quantization as a tool for representation theory \cite{Kirillov} and the relations between $G^{ \mathbb{C} } $ and $ G$ are discussed for example in \cite{BH}. The interplay between the theory of co-adjoint orbits as a tool of representation theory, 
   nilpotent orbits as algebraic varieties arising in supersymmetric  gauge theories, and harmonic analysis on homogeneous spaces (studied 
    in Kaluza-Klein reductions in physics) promises to be a fruitful area for future investigations.

It is also  interesting to note that co-adjoint orbits have played a role in connection with emergent geometry in Matrix theory \cite{BFSS,BKV0511}. For example, finite matrix approximations of $SO(2k)/U(k) ,  SO(2k+1)/U(k) , SO(2k)/ (U(k-1) \times U(1))  $ play a role 
in connection with higher dimensional fuzzy spherical branes \cite{HoRam02,fuzzyodd,Kimura0301}. Mathematical applications and physical interpretation of   finite matrix approximations in the context of the hyperK\"ahler moduli spaces under study here is a very interesting avenue for the future.

\section*{Acknowledgements}

A.~H.~ is supported by 
STFC Consolidated Grant ST/J0003533/1, and
EPSRC Programme Grant EP/K034456/1.
S.~R.~  is supported by  by STFC Grant ST/J000469/1, String Theory, Gauge Theory, and Duality.
D.~R-G.~ is partly supported by the spanish grant MINECO-13-FPA2012-35043-C02-02, the Ramon y Cajal grant RYC-2011-07593 as well as the EU CIG grant UE-14-GT5LD2013-618459.
A.~H.~ would like to thank Bo Feng and Rijun Huang for their very kind hospitality at the university of Zhejiang in Hangzhou where the final stages of this paper were completed.
D.~R-G.~ would like to greatly acknowledge the hospitality of the Center For String Theory at Queen Mary where this work was initiated. The authors would like to thank G.Ferlito, Y.Hui-He, V.Jejjala, R.Kalveks and M.Sperling for useful comments and discussions which led to the present paper.

\end{document}